%% file: main.tex
\documentclass[sigconf]{acmart}

\input{header.tex}

\copyrightyear{2023}
\acmYear{2023}
\setcopyright{acmlicensed}\acmConference[CIKM '23]{Proceedings of the 32nd ACM International Conference on Information and Knowledge Management}{October 21--25, 2023}{Birmingham, United Kingdom} \acmBooktitle{Proceedings of the 32nd ACM International Conference on Information and Knowledge Management (CIKM '23), October 21--25, 2023, Birmingham, United Kingdom}
\acmPrice{15.00}
\acmDOI{10.1145/3583780.3614855} \acmISBN{979-8-4007-0124-5/23/10}

\begin{document}

\title{Dual Intents Graph Modeling for User-centric Group Discovery}

\author{Xixi Wu}
  \email{21210240043@m.fudan.edu.cn}
\author{Yun Xiong}
\authornote{Corresponding author}
 \email{yunx@fudan.edu.cn}
 \author{Yao Zhang}
 \email{yaozhang@fudan.edu.cn}
  \affiliation{%
     \institution{Shanghai Key Laboratory of Data Science, School of Computer Science, Fudan University}
     \city{Shanghai}
    \country{China}
  }

\author{Yizhu Jiao}
  \email{yizhuj2@illinois.edu}
  \affiliation{%
    \institution{University of Illinois at Urbana-Champaign}
    \state{IL}
    \country{USA}
  }

\author{Jiawei Zhang}
  \email{jiawei@ifmlab.org}
  \affiliation{%
     \institution{IFM Lab, Department of Computer Science, University of California, Davis}
     \state{CA}
     \country{USA}
  }

\renewcommand{\shortauthors}{Wu, et al.}

\begin{abstract}

Online groups have become increasingly prevalent, providing users with space to share experiences and explore interests. Therefore, user-centric group discovery task, \ie, recommending groups to users can help both users' online experiences and platforms' long-term developments. Existing recommender methods can not deal with this task as modeling user-group participation into a bipartite graph overlooks their item-side interests. Although there exist a few works attempting to address this task, they still fall short in fully preserving the social context and ensuring effective interest representation learning.

In this paper, we focus on exploring the \textit{intents} that motivate users to participate in groups, which can be categorized into different types, like the \textit{social-intent} and the personal \textit{interest-intent}. The former refers to users joining a group affected by their social links, while the latter relates to users joining groups with like-minded people for interest-oriented self-enjoyment. To comprehend different intents, we propose a novel model, \textbf{DiRec}, that first models each intent separately and then fuses them together for predictions. Specifically, for social-intent, we introduce the hypergraph structure to model the relationship between groups and members. This allows for more comprehensive group information preservation, leading to a richer understanding of the social context. As for interest-intent, we employ novel structural refinement on the interactive graph to uncover more intricate user behaviors, item characteristics, and group interests, realizing better representation learning of interests. Furthermore, we also observe the intent overlapping in real-world scenarios and devise a novel self-supervised learning loss that encourages such alignment for final recommendations. Extensive experiments on three public datasets show the significant improvement of DiRec over the state-of-the-art methods.

\end{abstract}



\maketitle

\input{sec_1_intro.tex}

\input{sec_2_preliminary.tex}

\input{sec_3_method.tex}

\input{sec_4_exp.tex}

\input{sec_5_related_work}

\input{sec_6_conclusion.tex}

\section*{Acknowledgements}
This work is partially supported by the National Natural Science Foundation of China Projects No. U1936213, No.62206059, China Postdoctoral Science Foundation 2022M710747, and NSF through grants IIS-1763365 and IIS-2106972.



\balance{
\bibliographystyle{ACM-Reference-Format}
\bibliography{ref}
}

\end{document}

%% file: header.tex
\usepackage[labelformat=simple]{subcaption}
\usepackage{tabularx}

\usepackage{booktabs}  

\usepackage{graphicx}
\usepackage{bm}
\usepackage{url}
\usepackage{balance}

\usepackage{amsthm}
\usepackage{bbm}  
\usepackage{multirow}
\usepackage{algorithm}
\usepackage{algpseudocode}

\usepackage{enumitem}

\usepackage{hyperref}

\newcommand{\ie}[0]{\textit{i.e.}}
\newcommand{\eg}[0]{\textit{e.g.}}

%% file: sec_1_intro.tex
\section{Introduction}

\begin{figure}[!t]
    \centering
    \includegraphics[width=8cm]{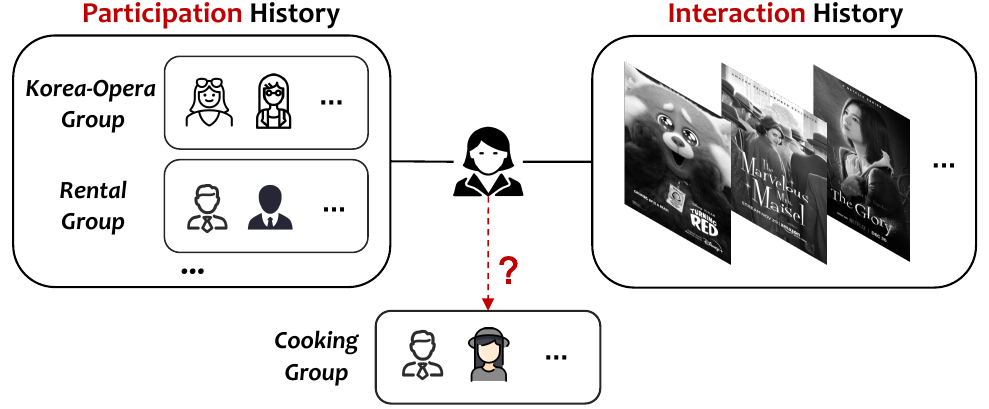}
    \caption{Illustration of User-centric Group Discovery (UGD) task on Douban platform. With group participation and item interactions, UGD aims to suggesting groups for users. }
    \label{fig:task}
\end{figure}

With the rise of social media, online groups have become increasingly prevalent, providing users with a space to share experiences, explore new interests, and establish social connections \cite{CFAG, ConsRec}. For instance, on Douban platform\footnote{https://www.douban.com/}, people who are following the same TV series often form groups to discuss the ongoing plot, fostering a sense of involvement with the show. Similarly, on Steam platform\footnote{https://store.steampowered.com/}, players also form groups to discuss about the gaming strategies and plan future play sessions. In terms of platforms, such users' attachment to groups can significantly boost the participation and retention rate of users. Therefore, user-centric group discovery (UGD) task, \ie, recommending groups to users, can help both users' online experiences and platforms' long-term developments, which, however, is still under-explored.

Existing recommender systems \cite{NCF, LightGCN, NGCF, SGL, DCCF, SimGCL} focus more on discovering relevant items for individual users. Typically, they build user-item bipartite graph based on observed interactions and then employ graph neural networks \cite{GIN, GAT, GraphSAGE, GCN} to propagate collaborative signals for recommendation. However, these approaches may not be appropriate for user-centric group discovery task. Groups contain rich social context and specific interests that cannot be fully captured by modeling interaction histories alone. Additionally, user-group participation can involve various motivations such as social links and personal hobbies, requiring a deeper understanding of users' \textit{intents} beyond their interactions. Thus, user-centric group discovery task differs from traditional recommendation settings, emphasizing the need for a unique approach that can effectively address these specific requirements.

\begin{figure}[!t]
    \centering
    \includegraphics[width=8.5cm]{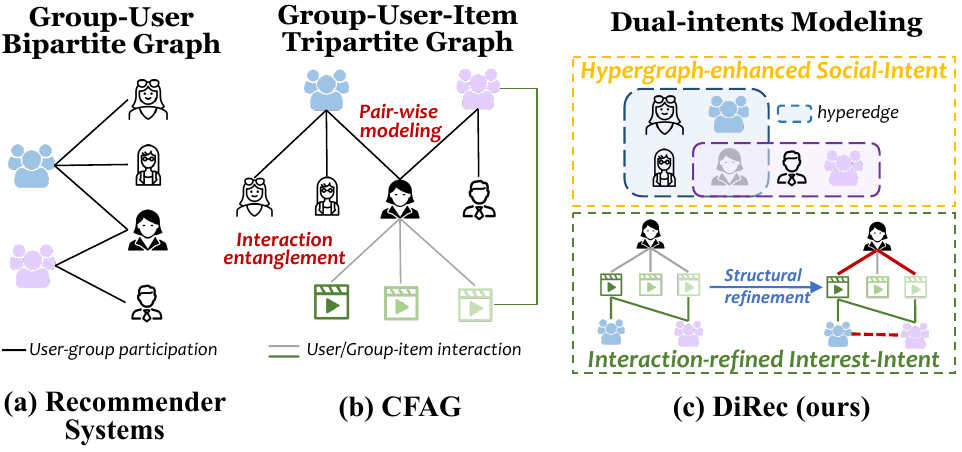}
    \caption{Modeling Method Comparison. (a) Illustration of adopting recommender systems \cite{LightGCN, NGCF, SimGCL} to UGD task. Regarding group as an item and employing recommendation methods to UGD is not ideal as missing interest-side information. (b) CFAG \cite{CFAG} unifies all behavioral data into a tripartite graph and thus has two drawbacks: i) pair-wise modeling of group-member relations leads to incomplete social context preservation; ii) entangled interaction data causes indistinguishable interest learning. (c) Our DiRec proposes a dual intents modeling method including hypergraph-enhanced social-intent and interaction-refined interest-intent.}
    \label{fig:comp}
    \vspace{-2.0em}
\end{figure}

Although prior work CFAG \cite{CFAG} has explored the integration of user-participation and item-interactions for group discovery, it still has two notable drawbacks: i) \textbf{incomplete social context preservation}. As shown in Figure \ref{fig:comp}, CFAG splits the comprehensive \textit{tuple-wise} group-member relations into multiple \textit{pair-wise} connections between users and groups, leading to group-side information loss and hindering further social signals propagation; ii) \textbf{indistinguishable interests representation}. Simply unifying all behavioral data among users, items, groups into a tripartite graph for representation learning  may cause different semantics and diverse preference messages to entangle with each other, with propagating deeper layers, ultimately leads to indistinguishable representations \cite{HeteroGNN}. Therefore, a more nuanced approach is needed for the UGD task.

In this paper, we focus on exploring the underlying intents that drive users to participate in groups to improve recommendation. Users' intents can generally be categorized into \textit{social-intent} and personal \textit{interest-intent}. The former refers to users joining a group due to social impacts from online friends that are connected via social links. For example, users may join family groups to keep in touch with relatives or connect with friends. Meanwhile, the interest-intent motivates users to join groups with other members who have no social connections but share common interests instead, allowing them to engage and discuss their passions with like-minded individuals. For instance, an ardent Korean-Opera fan may join a Korean-Opera group to discuss the series with other enthusiasts for further enjoyment.

To better understand aforementioned intents, we propose a novel model, \textbf{DiRec}, that first models each intent independently and then fuses them together for final predictions. Our approach to intents learning is characterized by two key innovations: i) \textbf{Hypergraph-enhanced Social-Intent Modeling.} To fully capture the tuple-wise relationships between groups and corresponding members, we propose a hypergraph structure for modeling by connecting each group node to corresponding members' nodes within a hyperedge. Based on such expressive group representation, more informative social messages can be generated and then propagated, reinforcing both users' and groups' social-side features. In a nutshell, hypergraph modeling enables more comprehensive relationship preservation and more effective representation learning, thereby leading to a richer understanding of social context. ii) \textbf{Interaction-refined Interest-Intent Modeling.} As to interest-intent, we mainly mine both users' and groups' interests from their interaction histories. However, the raw interaction data can be either noisy \cite{LDenoise, Denoise, GTN, GraphTrendFiltering} or sparse \cite{DualGraph}, hampering the ability to accurately estimate  their preferences. Therefore, we conduct structural refinement on the interactive graph to uncover more intricate user behaviors, item characteristics, and group interests. Through this process, more implicit interactive patterns can be captured, allowing for better interests learning.

Though we separately model users' social-intent and interest-intent to capture each intent's distinction, in real-world scenarios, it is common for these intents to overlap. For example, a group of colleagues who work together in the same industry may primarily join a group for social intent, such as staying connected outside of work, while also having an interest in exploring the latest innovations in their field. This overlapping of social and interest intents highlights the need to consider their alignment. Therefore, we propose a novel self-supervised loss \cite{InfoNCE, SSLCV, TwinSSL} that encourages possible intent alignment, allowing dual intents fusion for recommendations.

To summarize, our contributions are as follows: 

\begin{itemize}[leftmargin=*, topsep=2pt]
    \item In this paper, we focus on the user-centric group discovery task and propose a novel model DiRec. Specifically, we uncover the dual intents behind user participation, \ie, social-intent and interest-intent, for making recommendations.
    \item For social-intent, we introduce the hypergraph structure for tuple-wise group-member relationship modeling, surpassing the limited expressiveness of existing bipartite graph modeling. As for interest-intent, we employ novel structural refinement on the interactive graph to uncover more intricate user behaviors and group interests, allowing better representation of interests.
    \item To account for the possible overlaps of dual intents, we propose a novel contrastive learning objective.
   
    \item We evaluate DiRec on three public datasets and show significant improvements over existing methods.  
\end{itemize}

%% file: sec_2_preliminary.tex
\section{Notation Preliminaries}

In this section, we first present the concept of hypergraph and then give a formal definition of user-centric group discovery (UGD) task to facilitate comprehension. Formally, we use bold capital letters (\eg, $\mathbf{X}$) and bold lowercase letters (\eg, $\mathbf{x}$) to represent matrices and vectors, respectively. We utilize non-bold letters (\eg, $x$) to denote scalars, and calligraphic letters (\eg, $\mathcal{X}$) to denote sets. Notations are summarized in Table \ref{tab:notation}.

Different from simple graphs, a hypergraph is a more general topological structure where the edge (namely hyperedge in the hypergraph) could connect two or more nodes \cite{HG}. Formally, a hypergraph can be represented as $HG = (\mathcal{V}, \mathcal{E}, \mathbf{H})$ where $\mathcal{V}$ is the vertex set, $\mathcal{E}$ is the hyperedge set, and $\mathbf{H} = [h_{ve}] \in \mathbb{R}^{|\mathcal{V}| \times |\mathcal{E}|}$ depicts the connectivity of the hypergraph as $h_{ve}=1$ if the hyperedge $e$ connects the vertex $v$, otherwise $h_{ve}$ = 0. For the preservation of group-user affiliations, hypergraph modeling is superior to previous bipartite graph modeling where \textit{tuple-wise} relationships are split into multiple \textit{pair-wise} connections. By connecting each group and its members within a hyperedge, the comprehensive social context can be preserved, benefiting further representation learning stage.

Specifically, let $\mathcal{U}= \{ u_1, u_2, ..., u_M \}, \mathcal{I}= \{ i_1, i_2, ..., i_N \},$ and $\mathcal{G} = \{ g_1, g_2, ..., g_K \}$ be the sets users, items, and groups, respectively, where $M, N,$ and $K$ are the sizes of these three sets. There are three types of observed interactions among $\mathcal{U}, \mathcal{I},$ and $\mathcal{G}$, namely, user-group participation, user-item interactions, and group-item interactions. We use $\mathbf{X} = [x_{jt}] \in \mathbb{R}^{M \times K}$ to represent the user-group participation, where $x_{jt}=1$ if user $u_j$ is a member of group $g_t$, otherwise $x_{jt}=0$. Analogously, we define user-item and group-item interactive matrices as $\mathbf{Y} \in \mathbb{R}^{M \times N}$ and $\mathbf{Z} \in \mathbb{R}^{K \times N}$, respectively. The element $y_{jk}=1$ if user $u_j$ has interacted with item $i_k$, otherwise $y_{jk}=0$. For the \textbf{user-centric group discovery task} \cite{CFAG} studied in this paper, its target is to conduct group recommendation, which is defined as follows:

\begin{definition}[\textbf{User-centric Group Discovery}]
    Given user, item, and group sets $\mathcal{U}, \mathcal{I},$ and $\mathcal{G}$, and their interactive matrices $\mathbf{X}, \mathbf{Y},$ and $\mathbf{Z}$, the user-centric group discovery (UGD) for a user $u \in \mathcal{U}$ is to predict a ranking list of groups $\{ g_l, g_m, ..., g_n \} \subset \mathcal{G}$ where user $u$ has no existing interactions with such groups yet but will be highly likely to be interested in them in the future.
\end{definition}

%% file: sec_3_method.tex
\section{Methodology}

\begin{figure*}[!t]
    \centering
    \includegraphics[width=17cm]{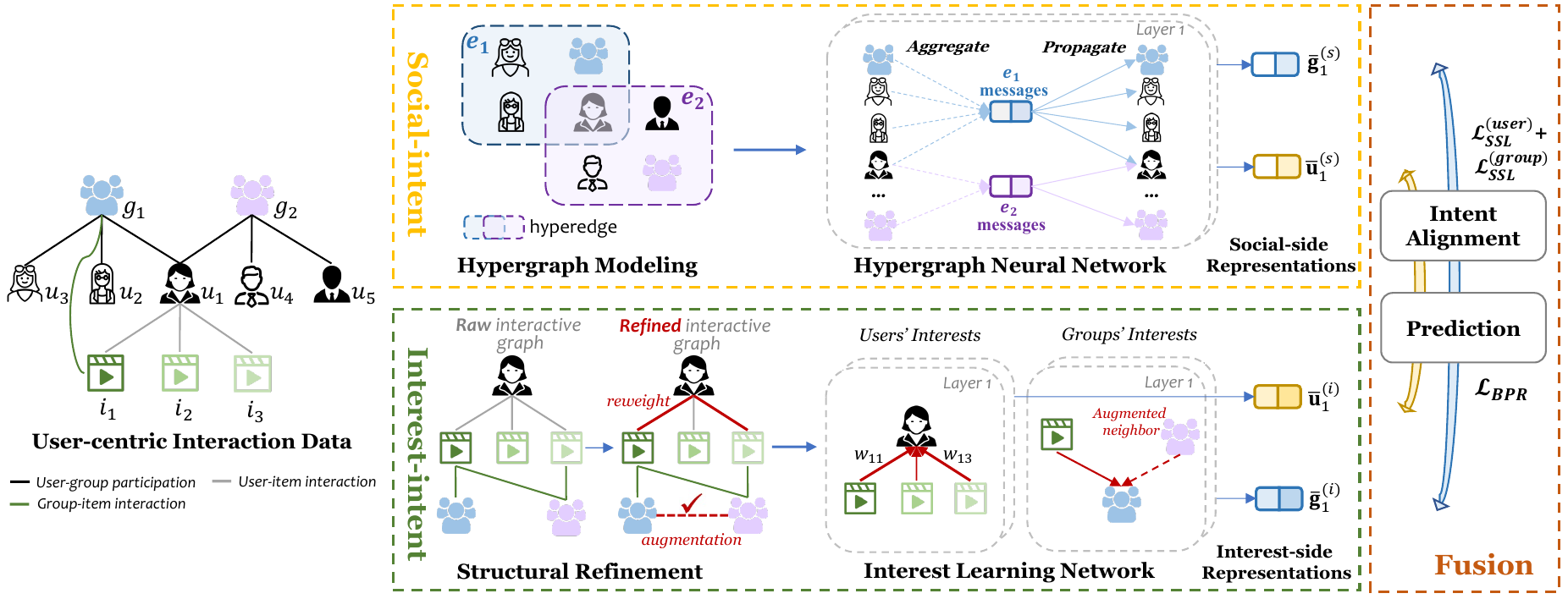}
    \caption{DiRec overview. In the left part, we present the interactive data between users, groups, and items (omit some group-item edges to avoid link interference). We model users' inclination towards groups from dual intents, including social-intent and interest-intent. For social-intent, we leverage hyperegraph for relationship preservation and employ hypergraph neural networks for representation learning. As to interest-intent, we conduct structural refinement to uncover intricate user behavior patterns and item characteristics, leading to better interests learning. Finally, dual intents are fused for optimization.}
    \label{fig:overview}
    
\end{figure*}

In this section, we present our proposed DiRec for the user-centric group discovery task. The overall framework is shown in Figure \ref{fig:overview}. We start by introducing the embedding layers that will be trained within this framework in Section \ref{sec:emb}. Next, we detail our approach for modeling and learning dual intents in Sections \ref{sec:social} and \ref{sec:interest}, respectively. Specifically, for social-intent, we leverage hypergraph for improved relationship preservation. As for interest-intent, we perform structural refinement to better uncover both users' and groups' interests. Finally, in Sections \ref{sec:combine} and \ref{sec:opt}, we demonstrate the fusion of dual intents for recommendation.

\begin{table}[!t]
    \centering
    \small
    \caption{Important Notations}
    \begin{tabular}{c|c}
       \toprule
       \textbf{Symbol}  & \textbf{Description} \\
       \midrule
       $\mathcal{U}, \mathcal{I}, \mathcal{G}$  &  Sets of users, items, and groups \\

       $M, N, K$ & Numbers of users, items, and groups \\

       \multirow{2}{*}{$\mathbf{X}, \mathbf{Y}, \mathbf{Z}$} & User-group, user-item, \\
         &  and group-item interactive matrices \\

        \midrule

      $\mathcal{M}_t=\{ u_{1}^{t}, ..., u_{|\mathcal{M}_t|}^t \}$ & Member set of group $g_t$ \\ 

       $\mathcal{I}_j = \{ i_1^j, ..., i_{|\mathcal{I}_j|}^j \}$ & Interaction set of user/group $j$ \\

       $\mathcal{N}_k=\{ u_{1}^{k}, ..., u_{|\mathcal{N}_k|}^k \}$ & Neighboring set of item $i_k$ in $G^{(u)}$ \\ 
       \midrule

       $ \mathbf{U} = [\mathbf{U}^{(s)} \| \mathbf{U}^{(i)}]$ & Embedding table of users \\
       $ \mathbf{G} = [\mathbf{G}^{(s)} \| \mathbf{G}^{(i)}] $ & Embedding table of groups \\
       $\mathbf{I}$ & Embedding table of items \\ 

       \midrule

       $HG^{(s)} = (\mathcal{V}^{(s)}, \mathcal{E}^{(s)}, \mathbf{H}^{(s)})$ & Social hypergraph \\ 

       $G^{(u)}=(\mathcal{V}^{(u)}, \mathcal{E}^{(u)}, \mathbf{A}^{(u)})$ & User interactive graph \\ 
       $G^{(g)}=(\mathcal{V}^{(g)}, \mathcal{E}^{(g)}, \mathbf{A}^{(g)})$ & Group interactive graph \\ 

       \midrule 

       $\mathbf{\overline{U}} = [ \mathbf{\overline{U}}^{(s)} \| \mathbf{\overline{U}}^{(i)}]$ & Refined users' embedding \\ 

       $\mathbf{\overline{G}} = [ \mathbf{\overline{G}}^{(s)} \| \mathbf{\overline{G}}^{(i)}]$ & Refined groups' embedding \\

       \bottomrule
       
    \end{tabular}
    \label{tab:notation}
    \vspace{-1.0em}
\end{table}

\subsection{Embedding Layer}\label{sec:emb}

We maintain three embedding tables $\mathbf{U}, \mathbf{I}$, and $\mathbf{G}$ to represent the trainable embeddings for users, items, and groups, respectively. Given that we infer a user's inclination towards groups via dual intents, we consider each user and group containing dual information, namely, social-side and interest-side representation. To extract a user $u_j$'s dual information from raw embedding $\mathbf{u}_j = \mathbf{U}_{j,:}$, we apply the following procedure:

\begin{equation*}
   \mathbf{u}_{j}^{(s)}, \mathbf{u}_{j}^{(i)} = \textbf{ExtractLayer}(\mathbf{u}_j),
\end{equation*}
\noindent where $\mathbf{u}_{j}^{(s)}$ and $\mathbf{u}_j^{(i)}$ represent user $u_j$'s social-side and interest-side representation, respectively. Similarly, we obtain group $g_t$'s social-side and interest-side embedding as $\mathbf{g}^{(s)}_t, \mathbf{g}^{(i)}_t = \textbf{ExtractLayer}(\mathbf{g}_t)$, where $\mathbf{g}_t = \mathbf{G}_{t,:}$ corresponds to the raw embedding of group $g_t$.

There are different ways to implement $\text{ExtractLayer}(\cdot)$. For instance, it can be linear transformation, \ie, $\mathbf{u}_j^{(s)} = \mathbf{W}^{(s)} \mathbf{u}_j $, $ \mathbf{u}_j^{(i)} = \mathbf{W}^{(i)} \mathbf{u}_j$, where $\mathbf{W}^{(s)}$ and $\mathbf{W}^{(i)}$ are two trainable weight parameter matrices. Moreover, we can also simply define $\text{ExtractLayer}(\cdot)$ as a function that splits a vector into two subvector components, \ie, $[ \mathbf{u}_j^{(s)} \| \mathbf{u}_j^{(i)}] = \mathbf{u}_j$ where $\|$ is the concatenation operation. Both function definition will work for our method, and, by default, we opt for the latter approach due to its simplicity and effectiveness. The shape of item embedding matrix $\mathbf{I}$ is $\mathbb{R}^{N \times d}$ while the shapes of $\mathbf{U}$ and $\mathbf{G}$ are $\mathbb{R}^{M \times 2d}$ and $\mathbb{R}^{K \times 2d}$ ($d$ is the embedding dimension), respectively.

\subsection{Social-Intent Learning}\label{sec:social}
In this subsection, we elaborate on social-intent learning process. To be more specific, we introduce a novel hypergraph structure for modeling tuple-wise relationships between groups and corresponding members. Through hypergraph neural networks, we are able to enhance the social-side representations of both users and groups by fully preserving their social context.

\subsubsection{Social Hypergraph Construction}
We construct a social hypergraph $HG^{(s)}=(\mathcal{V}^{(s)}, \mathcal{E}^{(s)}, \mathbf{H}^{(s)})$ for modeling existing user-group participation. Specifically, the nodes within this hypergraph contain all groups and users as $\mathcal{V}^{(s)}=\mathcal{U} \cup \mathcal{G}$, and each group node and their members' nodes are connected via a unique hyperedge as $\mathcal{E}^{(s)} = \mathcal{G}$. And $\mathbf{H}^{(s)} = [h_{vt}^{(s)}] \in \mathbb{R}^{(M+K) \times K}$ is the social hypergraph adjacency matrix as $h^{(s)}_{vt}=1$ if node $v \in \mathcal{M}_t \cup \{ g_t \}$, where $\mathcal{M}_t$ is the member set of group $g_t$. For instance, as shown in Figure \ref{fig:overview}, group $g_1$ contains three members, therefore, nodes $g_1$, $u_1, u_2$, and $u_3$ are connected within the hyperedge $e_1$.

We would like to emphasize that hypergraph modeling is superior to previous \textbf{bipartite graph modeling} where tuple-wise relationships are split into multiple pair-wise connections. By connecting each group and its member within a hyperedge, the comprehensive social context of each group can be preserved \cite{HGMotif}, allowing for group-level social signal propagating, and, finally, benefiting the representation learning. Empirical study also shows the effectiveness of hypergraph modeling than bipartite graph modeling, which will be shown in Section \ref{sec:hg}.

\subsubsection{Social Context Representation Learning}

Based on the constructed social hypergraph, we further conduct representation learning to obtain refined users' and groups' social-side representations.

In hypergraph neural networks \cite{HGNN, ContrastiveHG, AllSet, UniGNN}, hyperedges serve as mediums for processing and transferring information. More specifically, each hyperedge first aggregates the information of its connected nodes to generate messages, which are then used to update nodes' representations \cite{ConsRec}. As previously mentioned, we connect each group and its members into a single hyperedge. During the message aggregation stage, the messages carried by each hyperedge are highly expressive, as incorporating both member compositional information and the group's distinct social representation. Therefore, both groups' and users' final social representations can be enhanced from these expressive hyperedge messages.

Formally, we feed the concatenation of users' and groups' social-side embeddings $[\mathbf{U}^{(s)} \| \; \mathbf{G}^{(s)}]$ as well as social hypergraph adjacency matrix $\mathbf{H}^{(s)}$ to hypergraph neural networks \cite{HGNN}. After propagating several layers, we obtain the final users' and groups' social-side representations as follows:

\begin{equation*}
    \mathbf{\overline{U}}^{(s)}, \mathbf{\overline{G}}^{(s)} = \textbf{HyperGNN}_{\Theta_1}([\mathbf{U}^{(s)} \| \; \mathbf{G}^{(s)}], \mathbf{H}^{(s)} ),
\end{equation*}

\noindent where $\Theta_1$ denotes the parameters of hypergraph neural networks. We employ the classical hypergraph neural network \cite{HGNN} due to its simple implementation and generally good performance. Its computation is that $\mathbf{X}^{(l+1)} = \sigma(\mathbf{D}_{\mathcal{V}}^{-\frac{1}{2}} \mathbf{H} \mathbf{D}_{\mathcal{E}}^{-1} \mathbf{H}^T \mathbf{D}_{\mathcal{V}}^{-\frac{1}{2}} \mathbf{X}^{(l)} \mathbf{\Theta}^{(l)} )$, where $\mathbf{X}^{(l)}$ denotes the features in the $l$-th layer, $\mathbf{\Theta}^{(l)}$ denotes the weight matrix in the $l$-th layer, and $\mathbf{D}_{\mathcal{V}}$ and $\mathbf{D}_{\mathcal{E}}$ denote the degree matrix of nodes and hyperedges, respectively. Although more sophisticated hypergraph neural networks exist \cite{AllSet, ContrastiveHG}, we leave their application as a future work since the focus of our paper is the overall design of dual intents.

\subsection{Interest-Intent Learning}\label{sec:interest}
Despite social aspect, users also join groups with like-minded people for personal enjoyment. Therefore, we propose the interest intent that involves mining interests from interaction histories. Specifically, we conduct structural refinement to better uncover their interests. However, dynamically learning structures may result in large parameter sizes, slow convergence, and unstable performance \cite{GSL}. These issues can hinder the scalability and extensibility of our framework. Therefore, we conduct structural refinement as a pre-computation stage, which can be easily implemented and does not impact overall efficiency.

\subsubsection{Structural Refinement on User-Item Interactions}
The challenges of distilling users' interest lie in noisy user-item interactions as they may contain certain noises that do not necessarily indicate his/her real personal preference, \eg, random click or surrogate shopping \cite{noise1, noise2, LDenoise, Denoise}. Therefore, it is necessary to differentiate between various interactive weights on the user-item interactive graph in order to better identify a user's true intention.

Our intuition is that, co-occurrence between items indicates collaborative signals, making it useful for capturing users' interests \cite{CFGNN}. For example, as shown in Figure \ref{fig:overview}, we find that in user-item interactive graph, items $i_1$ and $i_3$ are commonly consumed by some users, reflecting their similarities and user $u_1$'s interests. Consequently, we would assign more weights to interactions $u_1 - i_1$ and $u_1 - i_3$ to preserve such signal. Formally, we first measure the co-occurrence ratio between two items $i_k$ and $i_j$ via Salton Cosine Similarity (SC) \cite{SC} as $\textbf{SC}(k,j) = \frac{|\mathcal{N}_k \cap \mathcal{N}_j|}{\sqrt{|\mathcal{N}_k \cup \mathcal{N}_j|}}$ where $\mathcal{N}_k$ and $\mathcal{N}_j$ refer to the neighboring set of item $i_k$ and $i_j$, respectively, in the user-item interaction graph. Then, for re-assigning the interaction weight between user $u$ and item $i_j$, we would compute the normalized version of above similarities among user $u$'s all interacted items as $w_{uj} = \frac{1}{|\mathcal{I}_u|} \sum_{i_k \in \mathcal{I}_u} \textbf{SC}(k,j)$ where $\mathcal{I}_u$ denotes the interaction set of user $u$. Based on all observed interactions, we could obtain a new weight matrix $\mathbf{W}^{(u)} = [w_{uj}] \in \mathbb{R}^{M \times N}$. Finally, we refine the user-item interactive matrix with this newly computed matrix as $\mathbf{Y}' = \mathbf{Y} + \lambda_1 \mathbf{W}^{(u)}$ where $\mathbf{Y}'$ denotes the refined user-item interactive matrix and $\lambda_1$ represents the reweighting coefficient. Using $\mathbf{Y'}$, we construct refined user-item interactive bipartite graph $G^{(u)} = (\mathcal{V}^{(u)}, \mathcal{E}^{(u)}, \mathbf{A}^{(u)})$ where $\mathcal{V}^{(u)}=\mathcal{U} \cup \mathcal{I}, \mathcal{E}^{(u)} = \{ (u_j, i_k) | u_j \in \mathcal{U}, i_k \in \mathcal{I}, \text{and} \; y_{jk}=1 \}$, and $\mathbf{A}^{(u)} = \begin{bmatrix}\mathbf{0} \ \ \ \ \mathbf{Y}' \\ \mathbf{Y'}^{T} \ \ \mathbf{0} \end{bmatrix}$.

\subsubsection{Structural Refinement on Group-Item Interactions}
Different from noisy user-item interactions, group-item interactive records are extremely sparse \cite{GroupIM, CubeRec}, making it hard to estimate groups' interest-side representations with limited neighboring nodes on the interactive graph \cite{CFAG, DualGraph}. To overcome this limitation, we consider performing structural augmentation to enrich each group's neighborhood, thereby improving their interest learning with a larger number of neighbors.

In user recommendation scenarios, if two users have purchased the same item, they are likely to share similar interests \cite{NCF}. This idea of collaborative filtering can be extend to group-level: if two groups have once interacted with the same item, they may share similar preferences and can reinforce each other's representations toward similar interests. Therefore, on the group-item interactive graph, besides observed group-item links, we connect two groups if they share common interactions. In this way, we obtain a refined group-item graph $G^{(g)} = (\mathcal{V}^{(g)}, \mathcal{E}^{(g)})$ where $\mathcal{V}^{(g)} = \mathcal{G} \cup \mathcal{I}$, $\mathcal{E}^{(g)} = \{ (g_t, i_k) | g_t \in \mathcal{G}, i_k \in \mathbf{I}, \text{and} \; z_{tk}=1 \} \cup \{ (g_p, g_q) | g_p, g_q \in \mathcal{G}, |\mathcal{I}_p \cap \mathcal{I}_q| \geq 1 \}$ ($\mathcal{I}_p$ denotes the interaction set of group $g_p$). Based on augmented edges, we can obtain the adjacency matrix $\mathbf{A}^{(g)}$, which will be further leveraged for interests learning.

\subsubsection{Interests Learning}
Using both the adjusted user-item and group-item interactive graphs, we perform representation learning to obtain refined users' and groups' interest-side embeddings.

Specifically, for users' interests distilling, recall that we have re-assigned the interaction weight and thus obtain a new user interactive graph $G^{(u)}$, we would feed the concatenation of users' interest-side embeddings $\mathbf{U}^{(i)}$ and items' embeddings $\mathbf{I}$, together with the adjacency matrix $\mathbf{A}^{(u)}$ to graph neural networks \cite{NGCF, LightGCN} to refine users' final interests as follows:

\begin{equation*}
    \mathbf{\overline{U}}^{(i)} = \textbf{GNN}_{\Theta_2}([\mathbf{U}^{(i)} \| \; \mathbf{I}], \mathbf{A}^{(u)}),
\end{equation*}
\noindent where $\Theta_2$ denotes the trainable parameters of graph neural networks and $\mathbf{\overline{U}}^{(i)}$  represents users' final interest-side representations. In this way, users' interests can be effectively captured from most representative interactions, thus pinpointing his/her actual interests. Additionally, we calculate the interest-side representations of groups in a similar way:

\begin{equation*}
    \mathbf{\overline{G}}^{(i)} = \textbf{GNN}_{\Theta_3}([\mathbf{G}^{(i)} \| \; \mathbf{I}], \mathbf{A}^{(g)}),
\end{equation*}

\noindent where $\Theta_3$ denotes the trainable parameters and $\mathbf{A}^{(g)}$ denotes the adjacent matrix of refined group's interactive graph. Note that instead of unifying users, groups, items into a single graph, we separately deal with users' and groups' interaction data to better understand their respective interests. We employ \textbf{LightGCN} \cite{LightGCN} for the implementation of graph neural networks due to its simplicity and effectiveness.

\subsection{Combination of Dual Intents}\label{sec:combine}
Though we separately model social-intent and interest-intent for capturing each intent's distinction, we note that, in real-world scenarios, dual intents may have certain overlaps. To address such overlap, we propose a self-supervised learning objective to guide intent alignment.

More specifically, the crux of our intent alignment approach is to ensure that the user $u_j$'s social-side representation $\overline{\mathbf{u}}_j^{(s)}$ is more similar to his/her interest-side representation $\overline{\mathbf{u}}_j^{(i)}$ than other users' interest representations, \eg, $\overline{\mathbf{u}}_k^{(i)} \; \text{as} \; u_k \in \mathcal{U}$. Therefore, we employ the InfoNCE loss \cite{InfoNCE} for implementation as follows:

\begin{equation*}
    \mathcal{L}_{SSL}^{(user)_1} = - \sum_{u_j \in \mathcal{U}} \text{log} \frac{ \text{exp}(\text{sim}( \mathbf{\overline{u}}^{(s)}_j, \mathbf{\overline{u}}^{(i)}_j )) }{ \sum_{u_k \in \mathcal{U}}  \text{exp}(\text{sim}( \mathbf{\overline{u}}^{(s)}_j, \mathbf{\overline{u}}^{(i)}_k )) },
\end{equation*}

\noindent where $\text{sim}(\cdot)$ denotes a certain similarity measurement function, \ie, inner product. And this objective holds for interest-side representations as well:

\begin{equation*}
     \mathcal{L}_{SSL}^{(user)_2} = - \sum_{u_j \in \mathcal{U}} \text{log} \frac{ \text{exp}(\text{sim}( \mathbf{\overline{u}}^{(i)}_j, \mathbf{\overline{u}}^{(s)}_j )) }{ \sum_{u_k \in \mathcal{U}}  \text{exp}(\text{sim}( \mathbf{\overline{u}}^{(i)}_j, \mathbf{\overline{u}}^{(s)}_k )) }.
\end{equation*}

\noindent Finally, we compute user-side loss as $\mathcal{L}_{SSL}^{(user)} = \mathcal{L}_{SSL}^{(user)_1} + \mathcal{L}_{SSL}^{(user)_2}$. Additionally, we infer a group-side loss $\mathcal{L}_{SSL}^{(group)}$ in a similar manner. By imposing these self-supervised learning objectives, we enable the model to better capture overlaps in dual intents and to generate more informative representations for both users and groups.

\subsection{Prediction and Optimization}\label{sec:opt}

In this subsection, we show the detailed prediction and optimization process. After obtaining social-side and interest-side representations, we concate them to obtain final users' and groups' representations as follows:

\begin{equation*}
    \mathbf{\overline{U}} = [\mathbf{\overline{U}}^{(s)} \| \mathbf{\overline{U}}^{(i)}] \;,     \mathbf{\overline{G}} = [\mathbf{\overline{G}}^{(s)} \| \mathbf{\overline{G}}^{(i)}], 
\end{equation*}

\noindent where $\|$ denotes concatenation operation. Then for a user $u_j$ and group $g_t$, the ranking score of this user-group pair is calculated by the inner product: 

\begin{equation*}
    \hat{s}_{jt} = \mathbf{\overline{u}}_j \cdot \mathbf{\overline{g}}_t,
\end{equation*}

\noindent where $\mathbf{\overline{u}}_j = \mathbf{\overline{U}}_{j,:}$ and $\mathbf{\overline{g}}_t = \mathbf{\overline{G}}_{t,:}$ refer to $u_j$ and $g_t$ final representations, respectively. Then, we employ the Bayesian Personalized Ranking (BPR) loss \cite{BPR} for optimization as follows: 

\begin{equation*}
    \mathcal{L}_{BPR} = \sum_{(u_j, g_t, g_{t'}) \in \mathcal{D}} - \text{log} \; \sigma(\hat{s}_{jt} - \hat{s}_{jt'}) + \lambda_2 \| \Theta \|_2^2,
\end{equation*}

\noindent where $\mathcal{D} = \{(u_j,g_t,g_{t'})|  x_{jt=1} \; \text{and} \; x_{jt'}=0 \}$ is the training data with observed interactions and random sampled negative pairs. $\Theta$ denotes all trainable parameters within this framework, which is regularized by $\lambda_2$. Finally, we unify the ranking loss as well as self-supervised loss into a joint training framework as $\mathcal{L} = \mathcal{L}_{BPR} + \lambda_3 \mathcal{L}_{SSL}^{(user)} + \lambda_4 \mathcal{L}_{SSL}^{(group)}$ where $\lambda_3$ and $\lambda_4$ are two hyper-parameters.

%% file: sec_4_exp.tex
\section{Experiments}

In this section, we present our experimental setup and empirical results. Our experiments are designed to answer the following research questions (RQs):

\begin{itemize}[leftmargin=*, topsep=2pt]
    \item \textbf{RQ1 (Overall Performance):} How does DiRec perform compared with both representative recommendation methods and the start-of-the-art UGD method?
    \item \textbf{RQ2 (Ablation Study):} How do the key components of DiRec affect the model performance, \ie, social-intent, interest-intent, hypergraph modeling, structural refinement, and SSL loss?
    \item \textbf{RQ3 (Parameters Study):} How do different settings of hyper-parameters affect the model performance?
\end{itemize}

\subsection{Experimental Settings}

\subsubsection{Datasets}

Following prior work CFAG \cite{CFAG}, we conduct experiments on three real-world datasets: Mafengwo, Weeplaces, and Steam. The statistics of these datasets are listed in Table \ref{tab:dataset}. The Mafengwo and Weeplaces datasets comprise the user's travel history with a location-based social network, whereas Steam is a newly-released dataset, collected from the Steam online platform by CFAG. We found CFAG did not split the validation set for model selection, therefore, we \textbf{re-split} the dataset into train set, validation set, and test set with the ratios of 70\%, 10\%, and 20\%, respectively.

\begin{table}[!t]
    \centering
    \small
    \caption{Statistics of datasets}
    \begin{tabular}{c|c|c|c}
      \toprule
      \textbf{Dataset}   &  \textbf{Mafengwo} & \textbf{Weeplaces} & \textbf{Steam} \\
      \midrule
      \# Users & 1,269 & 1,501  & 11,099 \\
      \# Groups & 972 & 4,651 & 1,085 \\ 
      \# Items & 999 & 6,406  & 2,351 \\ 
      \# User-Group Participation & 5,574 & 12,258 & 57,654 \\ 
      \# User-Item Interactions & 8,676 & 43,942 & 444,776 \\ 
      \# Group-Item Interactions & 2,540 & 6,033 & 22,318 \\ 
      Avg. \# Groups/user & 4.39 & 8.17 & 5.19 \\ 
      Avg. \# Items/user & 6.84 & 29.28 & 40.07 \\ 
      Avg. \# Items/group & 2.61 & 1.29 & 20.57 \\ 
      \bottomrule
    \end{tabular}
    \label{tab:dataset}
\end{table}

\subsubsection{Baselines}

To show the effectiveness of our proposed DiRec, we compare it with the following representative baselines:

\noindent \textbf{Group Recommendation methods:} Group Recommendation (GR) task is defined as suggesting suitable items for groups \cite{ConsRec}. Therefore, GR methods focus on modeling groups' preferences, overlooking the inclination between users and groups. To empirically show the inferior performance of applying GR on UGD task, we consider the following GR methods. Note that to apply these methods to UGD task, we adapt them by replacing their initial group-item BPR loss with user-group BPR loss:
\begin{itemize}[leftmargin=*, topsep=2pt]
    \item \textbf{AGREE \cite{AGREE}} This is a classical group recommendation method that leverages attention mechanism to estimate group representations by adaptively aggregating members' representations.
    \item \textbf{ConsRec \cite{ConsRec}} This is the state-of-the-art group recommendation method that models group's preferences from member-level, item-level, and group-level views. 
\end{itemize}

\begin{table*}[!t]
    \centering
     \caption{UGD Performance comparison on three datasets with Recall (R) reported.}
    \begin{tabular}{c|c|ccc|ccc|ccc}
    \toprule
   \multirow{2}{*}{Method Type} &  Dataset   &  \multicolumn{3}{c|}{Mafengwo}   & \multicolumn{3}{c|}{Weeplaces}  & \multicolumn{3}{c}{Steam}   \\
    &   Metric  & R@5 & R@10 & R@20   & R@5 & R@10 & R@20   & R@5 & R@10 & R@20  \\    

       \midrule

   Group &   $\textbf{AGREE}_{\text{(SIGIR'18)}}$ & 0.1216 & 0.1716 & 0.2368 & 0.1680 & 0.2270 & 0.2911 & 0.1168 & 0.1768 & 0.2577 \\ 

     Recommendation & $\textbf{ConsRec}_{\text{(WWW'23)}}$ & 0.2204 & 0.3141 & 0.4063 & 0.2514 & 0.3472 & 0.4391 & \underline{0.1690} & 0.2440 & 0.3330 \\ 

       \midrule

    &   \textbf{MF-BPR} & 0.1832 & 0.2407 & 0.2973 & 0.1734 & 0.2278 & 0.2822 & 0.1207 & 0.1848 & 0.2671 \\ 

     &  $\textbf{NGCF}_{\text{(SIGIR'19)}}$ & 0.1999 & 0.2650 & 0.3284 & 0.1787 & 0.2392 & 0.2961 & 0.1330 & 0.2010 & 0.2903 \\ 

    Recommender &   $\textbf{LightGCN}_{\text{(SIGIR'20)}}$ & 0.2293 & 0.2930 & 0.3596 & 0.1791 & 0.2431 & 0.3069 & 0.1543 & 0.2343 & 0.3273 \\ 

    Systems &  $\textbf{SGL}_{\text{(SIGIR'21)}}$ & 0.2259 & 0.2956 & 0.3555 & 0.1810 & 0.2443 & 0.3046 & 0.1535 & 0.2336 & 0.3276 \\ 

    & $\textbf{SimGCL}_{\text{(SIGIR'22)}}$ & \underline{0.2309} & 0.2925 & 0.3585 & 0.1808 & 0.2477 & 0.3079 & 0.1544 & 0.2332 & 0.3271 \\ 

    & $\textbf{DCCF}_{\text{(SIGIR'23)}}$ & 0.2049 & 0.2649 & 0.3216 & 0.1858 & 0.2507 & 0.3171 & 0.1614 & 0.2417 & 0.3423 \\
 
      \midrule

   User-centric &   $\textbf{CFAG}_{\text{(WSDM'23)}}$ & 0.2274 & \underline{0.3242} & \underline{0.4194} & \underline{0.2855} & \underline{0.3824} & \underline{0.4893} & 0.1597 & \underline{0.2485} & \underline{0.3502} \\ 

 Group Discovery &    $\textbf{DiRec}_{\text{ours}}$ & \textbf{0.2653} & \textbf{0.3585} & \textbf{0.4549} & \textbf{0.3177} & \textbf{0.4119} & \textbf{0.4987} & \textbf{0.1702} & \textbf{0.2685} & \textbf{0.3708} \\


       \bottomrule
    \end{tabular}
    \label{tab:overall_r}
\end{table*}

\begin{table*}[!t]
    \centering
     \caption{UGD Performance comparison on three datasets with NDCG (N) reported.}
    \begin{tabular}{c|c|ccc|ccc|ccc}
    \toprule
  \multirow{2}{*}{ Method Type} &   Dataset   &  \multicolumn{3}{c|}{Mafengwo}   & \multicolumn{3}{c|}{Weeplaces}  & \multicolumn{3}{c}{Steam}   \\
    &   Metric  & N@5 & N@10 & N@20   & N@5 & N@10 & N@20   & N@5 & N@10 & N@20  \\    

       \midrule

   Group   &  $\textbf{AGREE}_{\text{(SIGIR'18)}}$ & 0.0887 & 0.1051 & 0.1218 & 0.1132 & 0.1330 & 0.1501 & 0.0742 & 0.0937 & 0.1143 \\ 
       
  Recommendation   &  $\textbf{ConsRec}_{\text{(WWW'23)}}$  & 0.1519 & 0.1832 & 0.2069 &  0.1714& 0.2037 & 0.2283 & 0.1034 & \underline{0.1335} & \underline{0.1604} \\ 

       \midrule

    &   \textbf{MF-BPR} & 0.1293 & 0.1484 & 0.1631 & 0.1170 & 0.1358 & 0.1507 & 0.0771 & 0.0979 & 0.1190 \\ 

   &   $\textbf{NGCF}_{\text{(SIGIR'19)}}$ & 0.1440 & 0.1655 & 0.1820 & 0.1213 & 0.1417 & 0.1572 & 0.0862 & 0.1082 & 0.1311 \\ 

  Recommender &   $\textbf{LightGCN}_{\text{(SIGIR'20)}}$ & 0.1713 & 0.1922 & 0.2093 & 0.1225 & 0.1443 & 0.1619 & 0.1013 & 0.1274 & 0.1513 \\ 

  Systems &  $\textbf{SGL}_{\text{(SIGIR'21)}}$ & 0.1718 & \underline{0.1945} & 0.2099 & 0.1235 & 0.1448 & 0.1614 & 0.1018 & 0.1278 & 0.1520 \\ 
   & $\textbf{SimGCL}_{\text{(SIGIR'22)}}$ & \underline{0.1725} & 0.1926 & 0.2095 & 0.1233 & 0.1460 & 0.1626 & 0.1015 & 0.1273 & 0.1514 \\ 

      & $\textbf{DCCF}_{\text{(SIGIR'23)}}$ & 0.1493 & 0.1691 & 0.1837 & 0.1256 & 0.1476 & 0.1659 & \underline{0.1053} & 0.1314 & 0.1573 \\

      \midrule

User-centric   &   $\textbf{CFAG}_{\text{(WSDM'23)}}$  & 0.1552 &  0.1867 & \underline{0.2111} & \underline{0.1938} & \underline{0.2264} & \underline{0.2551} & 0.1035 & 0.1324 & 0.1584 \\ 

  Group Discovery  &  $\textbf{DiRec}_{\text{ours}}$ & \textbf{0.1908} & \textbf{0.2208} & \textbf{0.2455} & \textbf{0.2246} & \textbf{0.2565} & \textbf{0.2797} & \textbf{0.1086} & \textbf{0.1405} & \textbf{0.1669} \\ 

       \bottomrule
    \end{tabular}
    \label{tab:overall_ndcg}
\end{table*}

\noindent \textbf{Recommender methods:} We also consider following item recommendation models to illustrate their infeasibility on the UGD task. As CFAG \cite{CFAG} does, we apply them on UGD task by treating each group as an item and thus only utilizing user-group interactions:
\begin{itemize}[leftmargin=*, topsep=2pt]
 \item \textbf{MF-BPR \cite{BPR}} This is a classical recommendation model that employs matrix factorization technique and is optimized by the pair-wise BPR loss.
 \item \textbf{NGCF \cite{NGCF}} This method is a variant of standard GCN \cite{GCN} that leverages high-order connectivity in a user-item bipartite graph.
 \item \textbf{LightGCN \cite{LightGCN}} This is a method based on NGCF with optimization in training efficiency and generation ability by removing feature transformation and non-linear activation.
 \item \textbf{SGL \cite{SGL}} This work performs contrastive learning on LightGCN model to augment node representations.
 \item \textbf{SimGCL \cite{SimGCL}} This is a strong GNN-based recommender model that incorporates an augmentation-free contrastive loss.
 \item \textbf{DCCF \cite{DCCF}} This is the state-of-the-art GNN-based recommender model with disentangled representations. 
\end{itemize}

\noindent \textbf{User-centric Group Discovery (UGD) methods:}
\begin{itemize}[leftmargin=*, topsep=2pt]
    \item \textbf{CFAG \cite{CFAG}} This is the only existing UGD method. It constructs a group-user-item tripartite graph and designs a specific tripartite graph convolution mechanism for representation learning.
\end{itemize}

\subsubsection{Evaluation Metrics} We evaluate UGD task by ranking the test groups with all non-interacted groups of users. We adopt Recall@\{5, 10, 20\} and NDCG@\{5, 10, 20\} as evaluation metrics. Higher Recall and NDCG indicate better performance. We evaluate the performance of all methods over the same metrics and test data.

\subsubsection{Implementation Details}
We implement DiRec in PyTorch. For all baselines, we utilize their official codes released in Github. We conduct all the experiments on GPU machines of Nvidia Tesla V100 with 32GB memory. We empirically set the weight of regularization loss $\lambda_2$ as 0.0001. To obtain the values of $\lambda_3$ and $\lambda_4$, we conduct the grid search over $\{0, 0.0001, 0.001, 0.01, 0.1, 1 \}$. All reported results are the average of \textbf{5 runs}. Our codes, data, and baseline implementations are available at \textcolor{blue}{\href{https://github.com/WxxShirley/CIKM2023DiRec}{https://github.com/WxxShirley/CIKM2023DiRec}}.

\subsection{Overall Performance (RQ1)}
We present the overall performance comparison in Tables \ref{tab:overall_r} and \ref{tab:overall_ndcg}. From these results, we note the following key observations:

\begin{itemize}[leftmargin=*, topsep=2pt]
  \item Our proposed DiRec achieves the best performance on all three datasets over all evaluation metrics, demonstrating its superiority. Compared with the start-of-the-art UGD model CFAG \cite{CFAG}, our model still realizes a significant improvement. Unlike CFAG, which unifies all types of nodes and interactive edges into a single graph, we preserve the distinction of dual intents (\ie, social- and interest- intent), resulting in better performance.
  
  \item Group Recommendation methods are not suitable for UGD task as both the classical method AGREE \cite{AGREE} and the state-of-the-art method ConsRec \cite{ConsRec} perform poorly on all datasets. The reason for this is that GR aims to estimate group interests accurately, failing to fully capture the inclination between users and groups.

  \item Existing recommender models also fail to address UGD tasks. Modeling observed user-group participation as a bipartite graph and conducting representation learning as NGCF \cite{NGCF}, LightGCN \cite{LightGCN}, SGL \cite{SGL}, SimGCL \cite{SimGCL}, and DCCF \cite{DCCF} only captures social-intent while interest-intent is overlooked. Their inferior performance highlights the necessity of dual intents modeling.
  
\end{itemize}

\subsection{Ablation Study (RQ2)}

In this subsection, we explore the effectiveness of key components within DiRec to demonstrate the rationale behind our proposal. Specifically, we evaluate the following components: dual intents modeling, the hypergraph modeling, the structural refinement, and the self-supervised learning loss. Due to space limitations, we only present the evaluation metrics of NDCG@\{10, 20\} in Table \ref{tab:ablation}, while the Recall shows the same trend.

\begin{table}[!t]
    \centering
    \small
     \caption{Ablation Study on three datasets with NDCG (N) reported. ``Social-'' and ``Interest-'' refer to the variant that only utilizes social-intent and interest-intent, respectively. ``w/o. HG'' denotes the variant that replaces hypergraph modeling with bipartite graph modeling. And ``w/o. UI Re.'' and ``w/o. GI Aug.'' refer to variants that eliminate reweighting and augmentation, respectively.}
    \begin{tabular}{c|cc|cc|cc}
    \toprule
      Dataset   &  \multicolumn{2}{c|}{Mafengwo}   & \multicolumn{2}{c|}{Weeplaces}  & \multicolumn{2}{c}{Steam}   \\
       Metric  & N@10 & N@20   & N@10 & N@20 & N@10 & N@20   \\    

       \midrule

     \textbf{Full} &  \textbf{0.2208} & \textbf{0.2455} & \textbf{0.2565} & \textbf{0.2797} & \textbf{0.1405} & \textbf{0.1669}  \\ 

     \midrule 

     \textbf{Social-} & 0.1954 & 0.2157 & 0.1593 & 0.1784 & 0.1373 & 0.1628 \\ 

     \textbf{Interest-} & 0.1561 & 0.1900 & 0.2426 & 0.2671 & 0.1097 & 0.1391 \\ 
     
       \midrule

       \textbf{w/o. HG} & 0.2110 & 0.2325 & 0.2490 & 0.2754 & 0.1253 & 0.1546 \\

       \textbf{w/o. UI Re.} & 0.2208 & 0.2453 & 0.2558 & 0.2788 & 0.1398 & 0.1664 \\

       \textbf{w/o. GI Aug.} & 0.2117 & 0.2393 & 0.2200 & 0.2417 & 0.1371 & 0.1639 \\

       \midrule

       \textbf{w/o. SSL} & 0.2194 & 0.2430 & 0.2525 & 0.2752 & 0.1395 & 0.1651 \\

       \bottomrule
    \end{tabular}
    \label{tab:ablation}
\end{table}

\subsubsection{Effectiveness of Dual Intents}
To showcase the effectiveness of dual intents modeling, we conduct an ablation study where we predict the ranking score \textbf{with only one intent}, \ie, either social-intent or interest-intent. The corresponding results are denoted as ``Social-'' and ``Interest-'' in Table \ref{tab:ablation}, respectively. It is obvious that utilizing only one intent leads to a significant degradation in performance, highlighting the necessity of dual intents modeling. Moreover, the importance of each intent varies across different datasets. For example, on Weeplaces, predicting only from the interest intent leads to better results, while on Steam and Mafengwo, social-intent plays a more crucial role.

\subsubsection{Effectiveness of Hypergraph Modeling}\label{sec:hg}
Recall that we construct a hypergraph to represent the relations between groups and user members, to evaluate its effectiveness, we replace the hypergraph with a user-group bipartite graph and then adopt LightGCN \cite{LightGCN} for representation learning. The results are denoted as ``w/o. HG'' in Table \ref{tab:ablation}. By comparing the results with the full DiRec model, we can see that hypergraph modeling leads to a significant performance improvement, while the bipartite graph approach is less effective in preserving the user-group participation.

\subsubsection{Effectiveness of Structural Refinement}
To handle the distinct properties of user-item and group-item interactions, we perform structural refinement on the raw interactive graph. Specifically, we adopt a re-weighting technique on noisy user-item interactions to extract the most representative interacted items, while using an augmentation method on sparse group-item interactions to learn better group characteristics. To conduct the ablation study, we exclude user-item re-weighting and group-item augmentation techniques individually, and the results are presented as ``w/o. UI Re.'' and ``w/o. GI Aug.'' in Table \ref{tab:ablation}, respectively.

\noindent \textbf{UI Reweight } By comparing the results of ``w/o. UI Re.'' and ``Full'' in Table \ref{tab:ablation}, we can observe that the introduction of UI reweighting leads to performance improvement. However, the improvement is relatively marginal for Mafengwo because the user-item interactions are already sparse and can reflect the true user preferences. On the other hand, for Weeplaces and Steam, where user-level interactions are abundant and may contain noise, performing reweighting results in a more significant improvement.

\begin{figure}[!t]
    \centering
    \includegraphics[width=8cm]{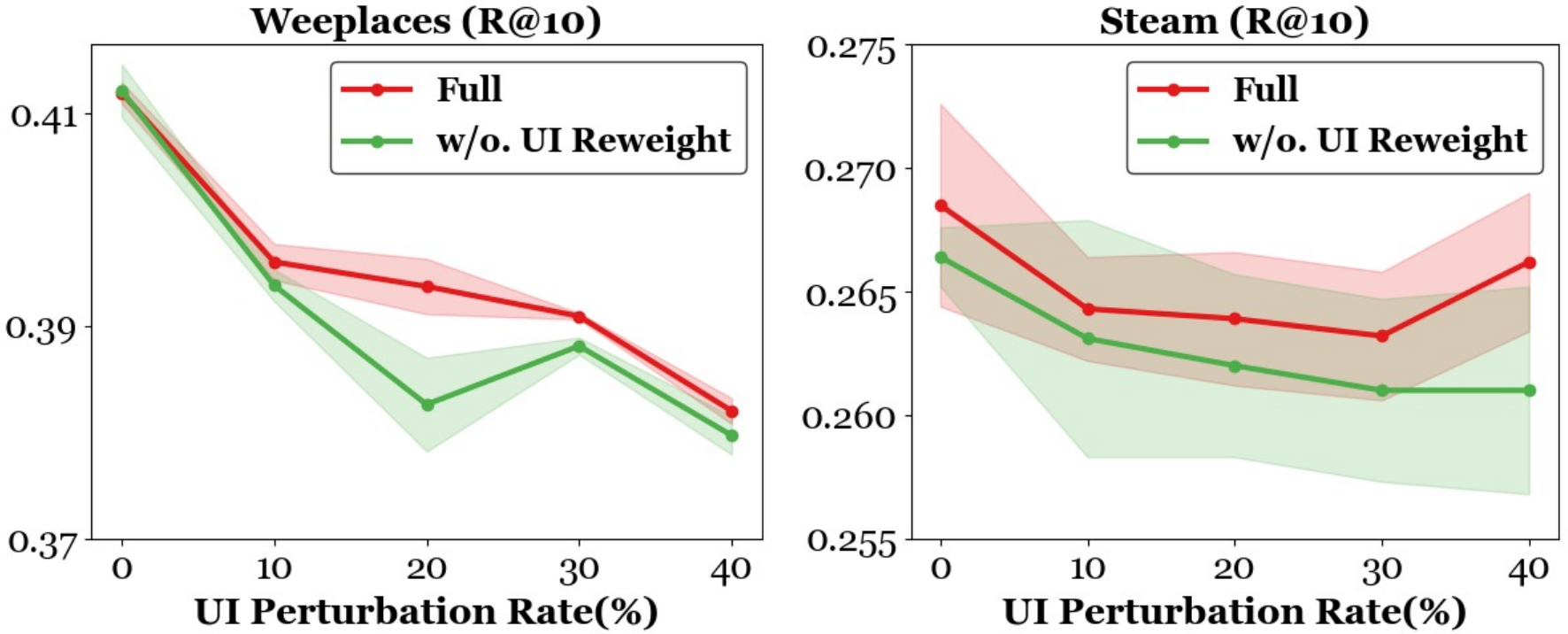}
    \caption{Performance comparison between DiRec and the variant the eliminates reweighting part under varying levels of user-item interaction perturbations.}
    \label{fig:perturbation}
    \vspace{-2.0em}
\end{figure}

To further investigate the effectiveness of introducing reweighting on user-item interactions, we perform structural perturbation on raw user-item interactions and then compare the performance of DiRec with its variant that removes the reweighting step. The results, as shown in Figure \ref{fig:perturbation}, reveal that when introducing higher levels of noise, the improvement brought by reweighting becomes more evident, demonstrating its effectiveness.

\noindent \textbf{GI Augmentation } By comparing the results of ``w/o. GI Aug.'' and ``Full'' in Table \ref{tab:ablation}, we can observe that performing augmentation on group-item interactive graph significantly enhance the overall performance. Notably, the improvement is most significant on Weeplaces, with an increase of 16.6\% on NDCG@10. This is because on Weeplaces, the group-item interactions are extremely sparse, as the average number of items that a group has interacted with is only 1.29. With so few neighbors, it is challenging to learn effective groups' interest-side representations. However, with the augmentation method, additional neighbors can help improve the target group's representations, leading to better performance.

\subsubsection{Effectiveness of Self-supervised Learning}

We also propose a self-supervised learning component that guides intent alignment for better representation. To evaluate its effectiveness, we exclude this loss and only optimize with BPR loss. This variant is denoted as ``w/o. SSL''. As shown in Table \ref{tab:ablation}, removing the self-supervised loss leads to a degradation in performance, indicating the importance of the self-supervised learning design.

\begin{figure}[!t]
    \centering
    \includegraphics[width=8cm]{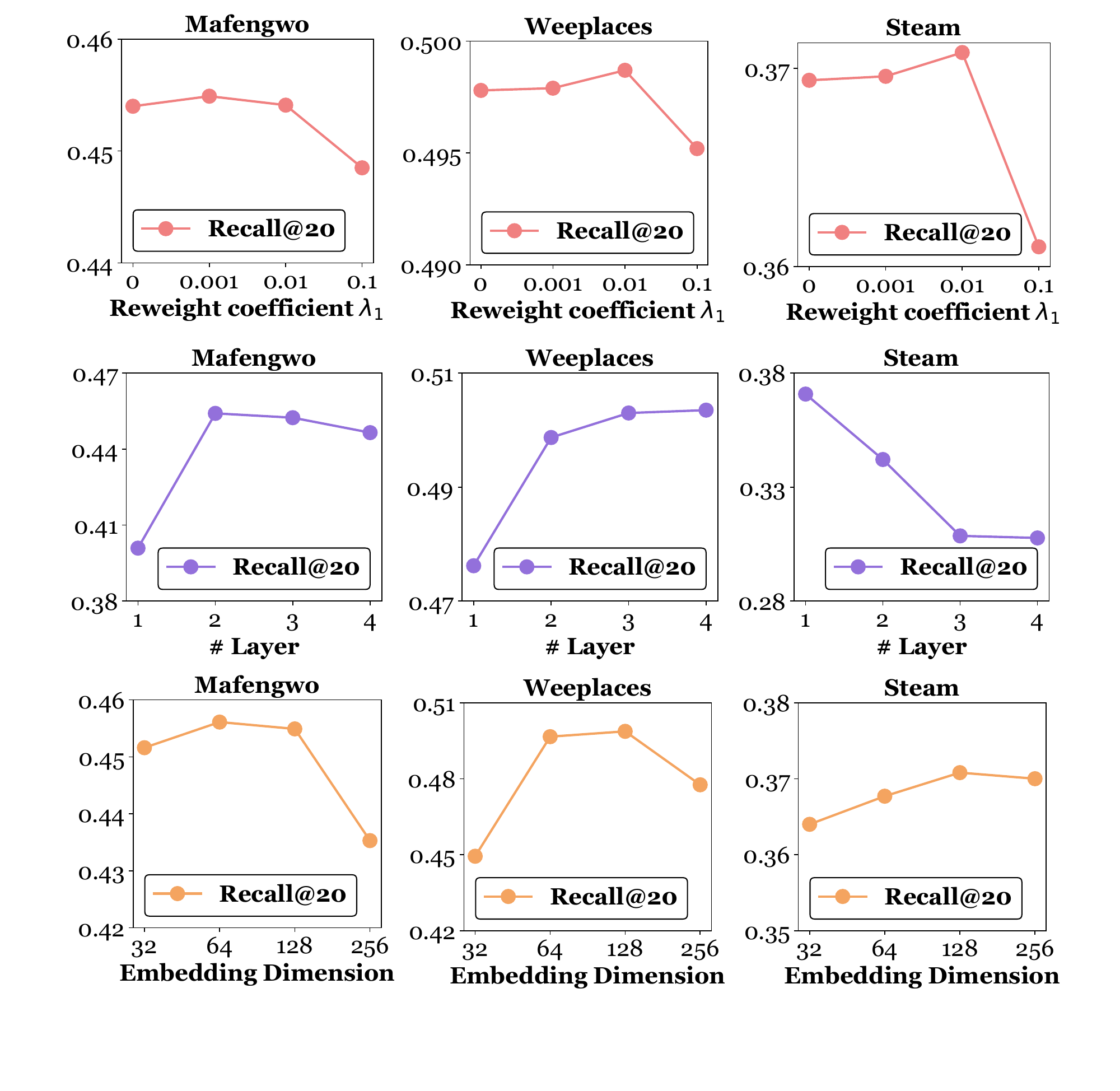}
    \caption{Parameters Study}
    \label{fig:param}
    \vspace{-2.0em}
\end{figure}

\subsection{Parameters Study (RQ3)}

In this subsection, we investigate the importance of key parameters within DiRec framework, \ie, the coefficient of user-item reweighting matrix $\lambda_1$, number of (hyper)graph convolutional layers $L$, and embedding dimension $d$. The results are shown in Figure \ref{fig:param}.

\subsubsection{The Coefficient of User-item Reweighting Matrix $\lambda_1$} We re-assign the weights of user-item interaction matrix as $\mathbf{Y}' = \mathbf{Y}+ \lambda_1 \mathbf{W}^{(u)}$ where $\lambda_1$ is the coefficient the controls the importance of newly computed interaction weights. To show its influence, we conduct a parameter study that tunes $\lambda_1$ across $\{ 0, 0.001, 0.01, 0.1\}$ and then compare performance difference. Figure \ref{fig:param} illustrates that with suitable $\lambda_1$ values, all three datasets experience performance improvement while overly large values of $\lambda_1$ may ruin the original interactions and lead to performance degradation. Therefore, for Mafengwo, we set $\lambda_1$ to 0.001, and 0.01 for Weeplaces and Steam.

\subsubsection{Number of Convolutional Layers $L$} We leverage both hypergraph neural networks \cite{HGNN} and graph neural networks \cite{LightGCN} for representation learning. Therefore, we perform the parameter study to show the influence of different numbers of convolutional layers. As Figure \ref{fig:param} indicates, we choose $L=2$ for Mafengwo and Weeplaces, while $L=1$ for Steam.

\subsubsection{Embedding Dimension $d$} We also study the influence of embedding dimensions within DiRec framework. From Figure \ref{fig:param}, we observe that the best performance is almost always achieved when the dimension is set to 128 across all datasets. Therefore, we choose 128 as the default setting.

%% file: sec_5_related_work.tex
\section{Related Works}

\subsection{UGD Related Works}

The task of user-centric group discovery was first proposed in CFAG \cite{CFAG}. As a pioneering work, CFAG unifies the diverse interactions between users, groups, and items into a tripartite graph and devises a tripartite graph convolutional network for representation learning and prediction. Although CFAG achieves ideal performance, we emphasize that its pair-wise modeling of group-member relations leads to incomplete social context preservation, and entangled interaction data causes indistinguishable interest learning.

Although CFAG is a prominent work in the field of UGD, no previous work has specifically targeted this task. However, there exist some closely related tasks such as (1) community detection, (2) group recommendation, and (3) recommender systems. Hence, we provide a brief introduction to these tasks and highlight the differences between them and UGD task.

\subsubsection{Community Detection}
Community detection task is defined as partitioning graph nodes into multiple groups, where internal nodes are more similar or more closely-related than the external \cite{Community1, Community2, CommunityGAN, vGraph, CLARE}. Therefore, with observed social links between users, community detection aims to identify new user groups. On the contrary, UGD task differs in the following ways: i) does not necessitate the user-user relations as input, ii) all group information is already known,  and iii) the objective is to predict users' ranking score towards a specific group.

\subsubsection{Group Recommendation}
Group recommendation task aims to suggesting suitable items for groups to boost group-level activities \cite{AGREE, ConsRec, GroupIM}. Existing group recommendation methods focus on aggregating diverse members' interests to estimate the group-level overall interest. Therefore, they mainly model both users and groups from interest-side, leaving social relations totally overlooked. 

Besides recommending items to groups, the term \textit{group recommendation} in literature also refers to recommending groups to their potential members. Traditional methods usually apply various algorithms to recover user-group membership matrices, using available side information such as semantic information from group descriptions in \cite{usergroup}, visual information from photos in \cite{usergroup2}, and user behaviors in different time periods \cite{usergroup3}. However, these methods require side information, which can lead to degraded performance when recommendation based solely on interaction information \cite{CFAG}. Thus, these methods are not suitable for UGD settings.

\subsubsection{Recommender Systems}

General recommendation models aim to recommend items for users based on observed user-item interactions. With graph neural networks (GNN)'s powerful capability of learning informative representations in graph data, GNN has been widely leveraged for recommendation scenarios \cite{NGCF, LightGCN, SGL, SimGCL}. Most previous works on Graph-based RS only focus on user-item bipartite graph \cite{LightGCN}. It is difficult to directly apply these works to UGD tasks with three relations to be managed: user-item, group-item and user-group relations.

\subsection{Intent-aware Recommendation}

In item recommendation models, the term ``intent'' typically refers to user's interests either for item contents or behavior purposes \cite{IntentCL}. To study such latent and implicit intentions behind users' behaviors, recent methods model intent as a set of fixed number of learnable variables and implement it in different ways. For instance, some methods derive intent via performing clustering on users' representations \cite{IntentCL} while others allow intent representations to be entirely learnable \cite{DCCF}. However, in our UGD task, we define ``intent'' as a user's inclination toward a group, and we differentiate intent into two aspects: social-intent and interest-intent.


%% file: sec_6_conclusion.tex
\section{Conclusion}
\label{sec:conclusion}

In this paper, we study the user-centric group discovery task and propose a novel model, DiRec. Specifically, we reveal the dual intents behind users to groups, \ie, social-intent and interest-intent. For social-intent, we employ hypergraph for relation preservation. To capture both users' and groups' item-level interests, we perform structural refinement on raw interactive graph. In future, we plan to study more fine-grained intents within social- and interest-intent to better understand users' behaviors. Besides, we plan to explore more complex intent alignment phenomenon and devise a more efficient self-supervised learning loss.